# Line positions and intensities of the $\nu_1$ band of $^{12}CH_3I$ using mid-infrared optical frequency comb Fourier transform spectroscopy

Adrian Hjáltén[a], Aleksandra Foltynowicz[a], Ibrahim Sadiek[a,b,]*

[a] *Department of Physics, Umeå University, 901 87 Umeå, Sweden*
[b] *Leibniz Institute for Plasma Science and Technology (INP), 17489 Greifswald, Germany*

* *Corresponding author. E-mail address: ibrahim.sadiek@inp-greifswald.de (I. Sadiek).*

**Abstract**

We present a new spectral analysis of the $\nu_1$ and $\nu_3+\nu_1-\nu_3$ bands of $^{12}CH_3I$ around 2971 cm$^{-1}$ based on a high-resolution spectrum spanning from 2800 cm$^{-1}$ to 3160 cm$^{-1}$, measured using an optical frequency comb Fourier transform spectrometer. From this spectrum, we previously assigned the $\nu_4$ and $\nu_3+\nu_4-\nu_3$ bands around 3060 cm$^{-1}$ using PGOPHER, and the line list was incorporated in the HITRAN database. Here, we treat the two fundamental bands, $\nu_1$ and $\nu_4$, together with the perturbing states, $2\nu_2+\nu_3$ and $\nu_2+2\nu_6^{\pm 2}$, as a four-level system connected via Coriolis and Fermi interactions. A similar four-level system is assumed to connect the two $\nu_3+\nu_1-\nu_3$ and $\nu_3+\nu_4-\nu_3$ hot bands, which appear due to the population of the low-lying $\nu_3$ state at room temperature, with the $2\nu_2+2\nu_3$ and $\nu_2+\nu_3+2\nu_6^{\pm 2}$ perturbing states. This spectroscopic treatment provides a good global agreement of the simulated spectra with experiment, and hence accurate line lists and band parameters of the four connected vibrational states in each system. It also allows revisiting the analysis of the $\nu_4$ and $\nu_3+\nu_4-\nu_3$ bands, which were previously treated as separate bands, not connected to their $\nu_1$ and $\nu_3+\nu_1-\nu_3$ counterparts. Overall, we assign 4665 transitions in the fundamental band system, with an average error of 0.00071 cm$^{-1}$, a factor of two better than earlier work on the $\nu_1$ band using conventional Fourier transform infrared spectroscopy. The $\nu_1$ band shows hyperfine splitting, resolvable for transitions with $J \leq 2 \times K$. Finally, the spectral intensities of 65 lines of the $\nu_1$ band and 7 lines of the $\nu_3+\nu_1-\nu_3$ band are reported for the first time using the Voigt line shape as a model in multispectral fitting. The reported line lists and intensities will serve as a reference for high-resolution molecular spectroscopic databases, and as a basis for line selection in future monitoring applications of CH$_3$I.

**1. Introduction**

Spectroscopic monitoring of methyl iodide, $CH_3I$ – a naturally occurring halogenated volatile organic compound – is essential for modelling the natural cycling of climate relevant trace gases as it is an important carrier of iodine from the oceans to the atmosphere [1]; and for avoiding personal exposure limits as it is toxic and being used in several industrial and agricultural applications [2]. In addition, $CH_3I$ has been proposed very recently together with other methylated halogens, e.g., $CH_3Br$, and $CH_3Cl$, as an exoplanetary capstone biosignature [3], i.e., a metabolic product with substantially lower false-positive potential, serving as confirmation for primary biosignatures such as $O_2$. For all of these applications, accurate spectroscopic models are needed for detection of $CH_3I$, particularly in the mid-infrared (mid-IR) region, where the absorption cross-section is larger compared to the near-infrared region.

In the mid-IR region, $^{12}CH_3I$ has been a subject of numerous studies using conventional Fourier transform spectroscopy (FT-IR) that provided spectroscopic information for the six fundamental bands: $\nu_1$, $\nu_2$, $\nu_3$, $\nu_4$, $\nu_5$, and $\nu_6$ [4-9] as well as some combination bands [8, 10]. A few of these studies reported line intensities, e.g., for the $\nu_6$ and the $2\nu_3$ bands [11-13]. The 3.3 μm region is dominated by the fundamental $\nu_1$ and $\nu_4$ bands originating from closely lying vibrations associated with the stretching of the C–H bonds of the molecule. In the normal mode analysis, the $\nu_1$ vibration is the symmetric (A1) stretch, and the $\nu_4$ vibration is the degenerate (E) stretch. The rovibrational lines of the $\nu_1$ band are more congested than the $\nu_4$ lines; however, the $\nu_1$ band is almost one order of magnitude stronger than the $\nu_4$ band, and hence it is promising for monitoring applications as a signature for $CH_3I$ in the different environments. The two previous studies of the $\nu_1$ band of $^{12}CH_3I$ by Paso et al. [7, 21] provided line positions, using an FT-IR spectrometer with a resolution of ~0.0054 cm$^{-1}$ (~160 MHz), which is larger than the Doppler width of $CH_3I$ at 296 K in this spectral range (FWHM: ~ 96 MHz). They also provided no information about the line intensities. In their spectral analysis, they accounted for the most striking perturbations, including the Fermi interactions with the $2\nu_2+\nu_3$ band, which shifts the band origin of the $\nu_1$ band by ~0.6 cm$^{-1}$ to higher wavenumber, and the two Coriolis interactions with the $\nu_2+2\nu_6^{\pm 2}$ band. However, in their spectroscopic analysis, they treated the bands separately and not in connection with their $\nu_4$ counterparts.

Quite recently, we presented high-resolution frequency comb-based Fourier transform spectroscopy (FTS) measurements of $^{12}CH_3I$ in the range from 2800 cm$^{-1}$ to 3160 cm$^{-1}$ [14]. Using these absorption spectra, we provided line lists for the $\nu_4$ band and the nearby $\nu_3+\nu_4-\nu_3$ hot band around 3060 cm$^{-1}$, and line intensities for 157 isolated transitions distributed over the entire $\nu_4$ band [14], and the data were incorporated into the recent HITRAN 2020 database [15]. Compared to conventional FT-IR spectroscopy based on incoherent light sources, the comb-based FTS allows acquisition times orders of magnitude shorter [16], which reduces the influence of changes in the sample during the measurement time. More importantly, by precisely matching the nominal resolution of the FTS to the repetition rate of the comb [17-19], the comb modes are sampled very accurately, which provides absolute calibration of the frequency scale and greatly reduced instrumental line shape. Finally, interleaving spectra measured at different repetition rates allows fully resolving the spectral features, even when their full width at half maximum (FWHM) is smaller than the spacing between the comb modes, e.g., in sub-Doppler spectroscopy [20].

The scope of this work is to extend the analysis of the high-resolution comb-based $^{12}CH_3I$ spectra used previously to the $\nu_1$ spectral region around 2971 cm$^{-1}$, and report for the first time line intensities of selected absorption lines of the $\nu_1$ band and the nearby $\nu_3+\nu_1-\nu_3$ hot band. The spectroscopic model is based on available microwave data and the earlier models of Paso et al. [7, 21] for the $\nu_1$ band and of Anttila et al. [6] for the $\nu_4$ band, however, the two fundamental $\nu_1$ and $\nu_4$ bands together with the perturbing states $2\nu_2+\nu_3$ and $\nu_2+2\nu_6^{\pm 2}$ are assumed to form a 'four-level system' interacting via a number of Coriolis and Fermi interactions. Similar spectroscopic treatment of the $\nu_1$ and $\nu_4$ states as interacting system was provided for $^{13}CH_3I$ by Alanko et al. [22]. The two hot bands in the vicinity of the $\nu_1$ and $\nu_4$ bands show up due to the population of the low-lying $\nu_3$ mode, located at 533 cm$^{-1}$, at room temperature. Series of hot bands $n\nu_4+\nu_6-n\nu_4$, with n = 1 to 5, were observed for $CH_2I_2$ [23], due to the population of the low-lying $\nu_4$ vibrations at room temperature. Similar series of hot bands have been observed very recently for $CH_2Br_2$ [24], which were previously misinterpreted in Ref. [25] as isotopic shifts of the different isotopologues of $CH_2Br_2$. The two hot bands of $^{12}CH_3I$, $\nu_3+\nu_1-\nu_3$ and $\nu_3+\nu_4-\nu_3$, are also assumed to form a four-level system together with the perturbing states: $2\nu_2+2\nu_3$ and $\nu_2+\nu_3+2\nu_6^{\pm 2}$. Such a spectroscopic model provides a good global agreement in the assigned transitions between the measurements and the simulations, and hence accurate band parameters for all the interacting vibrational states are reported. The broadband and the high-resolution capability of comb-based FTS together with the proposed spectroscopic model allowed us to create new line lists of the $\nu_1$ and $\nu_3+\nu_1-\nu_3$ bands. It also allowed us to revisit the analysis of the $\nu_4$ band [14], which was earlier treated separately and not coupled to the $\nu_1$ band as suggested here.

## 2. Experimental setup and procedures

The experimental setup and measurement procedure were described in detail in Ref. [14] and will be only briefly presented here. The setup consisted of a mid-IR frequency comb source, a multipass cell connected to a gas supply system, and an FTS, as shown schematically in Fig. 1. The comb source covered 360 cm$^{-1}$ around 2980 cm$^{-1}$ [26]. It was based on difference frequency generation from a common Yb:fiber oscillator, and was thus free from the carrier envelope offset frequency. The nominal repetition rate, $f_{rep}$, of the oscillator was 125 MHz and it was stabilized to a tunable direct digital synthesizer referenced to a GPS-disciplined Rb clock, by feeding back to a piezo-electric transducer (PZT) in the oscillator cavity.

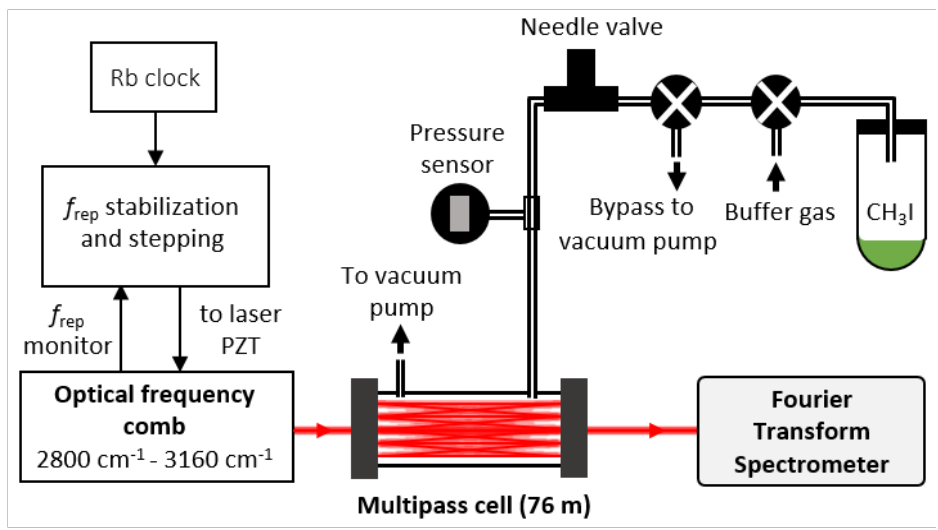

Fig. 1. Schematic of the experimental setup, including a stabilized frequency comb emitting in the mid-IR, a multipass cell, a sample supply system, and a Fourier transform spectrometer.

The comb beam passed through a Herriot-type multipass cell (Aerodyne AMAC-76LW) with a path length of 76 m. The cell was connected to a gas supply system and the pressure was measured using a pressure transducer (CERAVAC CTR 101 N) with 10 µbar resolution and specified relative accuracy of 0.15%. Analytical grade methyl iodide (Sigma-Aldrich 99%, containing copper as stabilizer) was contained in an Ace glass tube connected to the gas supply system and evaporated into the multipass cell through a needle valve. Prior to filling the cell, the headspace was evacuated several times through a bypass valve to eliminate impurities. For diluted samples, dry air was used as buffer gas, and all measurements were performed at room temperature (296.0 ± 0.6 K).

After the multipass cell, the beam was coupled to a home-built fast-scanning FTS with auto-balanced detection [27]. The optical path difference was calibrated by a continuous wave (CW) reference laser with wavelength, $\lambda_{ref}$ = 1563 nm. The interferograms were digitized on a PC and recorded using a home-written LabView$^{TM}$ program.

We recorded 13 sets of 200 interferograms at different repetition rate values separated by 15 Hz, corresponding to 11 MHz shift of the comb modes in the optical domain. The total acquisition time of all sets was 113 minutes. We averaged each set in the spectral domain after Fourier transforming, and normalized it to a reference spectrum consisting of 200 averages recorded at the first $f_{rep}$ step with the multipass cell evacuated. After normalization, we calculated the corresponding absorption spectra using the Lambert-Beer law and fitted a baseline consisting of a 4$^{th}$ order polynomial and a number of slowly varying sine terms to each spectrum while masking out the absorption lines. After subtracting the baseline, we interleaved the 13 spectra to obtain a final sampling point spacing of 11 MHz. For each of the measured 13 spectra, the sampling points were matched to the comb line positions by minimizing the instrumental line shape distortions, as explained in Refs [14, 17, 19]. We estimate that the optimization process contributes an uncertainty of 1.8 MHz to the line positions.

## 3. Results and discussions

### 3.1 High resolution spectra

Fig. 2 (a) presents the high-resolution spectrum of $^{12}CH_3I$ measured at a total cell pressure of 30 μbar in the region from 2900 cm$^{-1}$ to 3160 cm$^{-1}$, same as in Ref. [14]. The spectrum in this range is dominated by the fundamental $\nu_1$ band, centred at 2971 cm$^{-1}$, corresponding to the symmetric C–H stretch (CH$_3$ s-stretch), which appears as a parallel band with a doublet of P-branch/R-branch, and the fundamental $\nu_4$ band around 3060 cm$^{-1}$ corresponding to the degenerate asymmetric C–H stretch (CH$_3$ d-stretch), which appears as a perpendicular band with $Q_K(J)$ sub-bands. In our previous work [14], we presented the analysis of the $\nu_4$ band using the spectral window from 3010 to 3160 cm$^{-1}$; here, we focus on the $\nu_1$ band region from 2935 cm$^{-1}$ to 3010 cm$^{-1}$. Fig. 2 (b) shows a zoom on the $\nu_1$ band, which – similarly to the $\nu_4$ band – has resolved rovibrational lines, promising for future monitoring applications. This can be seen in panels (c) and (d) displaying segments of the P and R branches, respectively. We note that the central Q-branches as well as the strongest lines of the P and R-branches of the $\nu_1$ band are saturated. The spectrum shown here was used for the spectral simulations together with a spectrum measured at 110 μbar of CH$_3$I (not shown). The measurement at the higher pressure was used for the assignment of weak absorption lines in the spectrum.

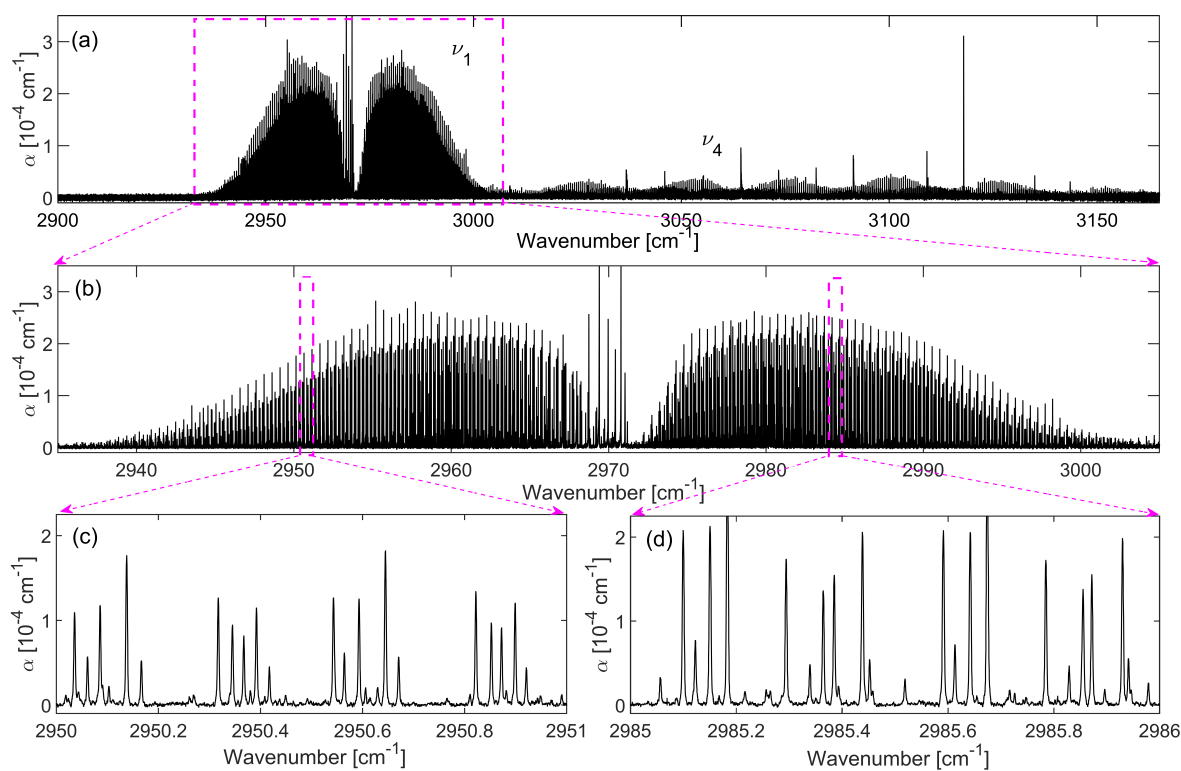

Fig. 2. (a) High-resolution absorption coefficient, α, of pure $^{12}CH_3I$ measured in the range from 2900 to 3160 cm$^{-1}$ at 30 μbar. Panel (b) shows the parallel $\nu_1$ band with its doublet of P- and R-branches. Panels (c) and (d) show zoomed-in spectra with the resolved rovibrational structures at the P and R branches of the $\nu_1$ band, respectively.

### 3.2. Spectral simulations and analysis

Methyl iodide is a prolate symmetric top molecule with an equilibrium structure of the $C_{3v}$ point group. Fig. 3 shows the optimized geometry of CH$_3$I with the principal axes system, where the yellow arrow represents the dipole moment change parallel to the principal c-axis due to the symmetric C–H stretch vibration of the $\nu_1$ band, and the blue arrows represent the displacement vectors of the vibration.

Mar 17, 2023

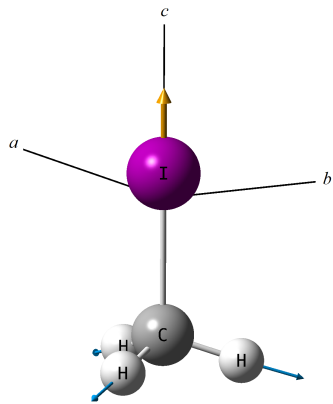

Fig. 3. The molecular structure of CH$_3$I. The yellow arrow represents the dipole moment change parallel to the principal c-axis because of the symmetric C–H stretch vibration, and the blue arrows represent the displacement vectors of the vibration.

The spectra were simulated and assigned with the help of PGOPHER [28]. The input parameters of the ground state, ν$_0$, were based on microwave measurements [29, 30], while those of the low-lying ν$_3$ state were based on FT-IR measurements [5]. The initial parameters of the upper ν$_1$ state were taken from the earlier simulations of Paso et al. [7]. The parameters of the ν$_4$ band were taken from our recent work [14].

As already outlined earlier, the ν$_1$ and the ν$_4$ vibrations are connected directly with each other via the Coriolis coupling, and indirectly with the two 2ν$_2$+ν$_3$ and ν$_2$+2ν$_6^{\pm2}$ perturbing states through a number of Fermi and Coriolis resonances. Therefore, a complete analysis should include all of these interactions simultaneously, taking into account that the two fundamental bands and the two perturbing states form a connected four-level system – the ν$_1$/ν$_4$ level system – as presented here.

Fig. 4(a) shows the interacting molecular vibrations in the fundamental, ν$_1$/ν$_4$ level system, including the most dominant Coriolis and Fermi interactions: (a) the Coriolis, C$_{4266}$, and Fermi, F$_{4266}$, resonances between the ν$_4$ band and the ν$_2$+2ν$_6^{\pm2}$ combination band; (b) the Coriolis resonance, C$_{4223}$, between the ν$_4$ band and the 2ν$_2$+ν$_3$ combination band, (c) the Fermi resonance, F$_{1223}$, between the ν$_1$ and the 2ν$_2$+ν$_3$ bands; (d) the Coriolis resonance, C$_{1266}$, between the ν$_1$ and the ν$_2$+2ν$_6^{\pm2}$ bands; and finally the Coriolis resonance, C$_{14}$, between the ν$_1$ and the ν$_4$ fundamental bands. The same molecular interactions are assumed for the hot band, ν$_3$+ν$_1$−ν$_3$/ν$_3$+ν$_4$−ν$_3$ level system, see Fig. 4(b), with the 2ν$_2$+2ν$_3$ and ν$_2$+ν$_3$+2ν$_6^{\pm2}$ bands as the perturbing states. The model adopted for the spectral analysis is similar to the models used by Paso et al. [7], and Anttila et al. [6] for the ν$_1$ and ν$_4$ bands of $^{12}$CH$_3$I, respectively.

The general form of the matrix elements of the Coriolis interactions is, for the case of ν$_4$ and the ν$_2$+2ν$_6^{\pm2}$ bands:

$$\langle v_4 = 1, l_4 = \pm 1, J, K |\widetilde{H}| v_2 = 1, v_6 = 2, l_6 = \pm 2, J, K \pm 1 \rangle = \pm [C_{4266} + C_{4266}^K + C_{4266}^J] \times \sqrt{J(J+1) - K(K \pm 1)} \quad (1),$$

and that of the Fermi interaction is

$$\langle v_4 = 1, l_4 = \pm 1, J, K |\widetilde{H}| v_2 = 1, v_6 = 2, l_6 = \mp 2, J, K \rangle = \pm [F_{4266} + C_{4266}^J] \times \sqrt{J(J+1))} \quad (2).$$

The *J*-dependent components have been included for all resonances shown in Fig. 4, while the *K*-dependent Coriolis parameters were put to zero.

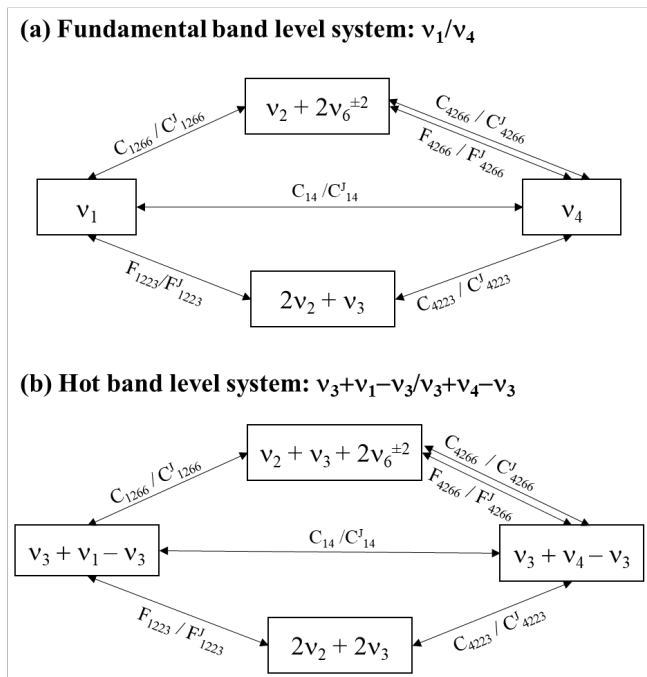

Fig. 4. Interacting vibrations included in the simulations of (a) the fundamental, $\nu_1/\nu_4$ level system, and (b) the hot band, $\nu_3+\nu_1-\nu_3/\nu_3+\nu_4-\nu_3$ level system.

Fig. 5(a) shows the measured absorption coefficient at 30 μbar of pure $CH_3I$ (black) together with the overall simulations in the range from 2935 cm$^{-1}$ to 3010 cm$^{-1}$. In this range, the simulated spectra, inverted for clarity, include the fundamental $\nu_1$ band (blue) and the nearby $\nu_3+\nu_1-\nu_3$ hot band (red). Note that the simulations of each system, i.e., the $\nu_1/\nu_4$ and $\nu_3+\nu_1-\nu_3/\nu_3+\nu_4-\nu_3$, involve four vibrational states together with their matrix elements as depicted in Fig. 4 (a and b). As shown in this figure, an overall very good match between the experiment and the simulations is obtained. The strong lines of the P and R branches have reduced intensities in the experimental spectrum because of absorption saturation, which can be seen by comparing it with the simulated spectra. Panel (b) shows an enlarged portion of the spectrum around 2969 cm$^{-1}$, where sub-bands of $Q_6(J)$ and $Q_7(J)$ are formed and degrading towards lower frequencies with increasing $J$ number. In this panel, weak rovibrational transitions (red) are observed and attributed to the $\nu_3+\nu_1-\nu_3$ hot band, as previously reported by Paso et al. [7, 21]. The strength of the $\nu_3+\nu_1-\nu_3$ hot band, as estimated from the intensities of the overlapping $Q_K(J)$ transitions, shown in Fig. 5(b), is ~7.5% of the fundamental $\nu_1$ band, which agrees with theoretical prediction assuming Boltzmann distribution.

The effect of the large $^{127}I$ nuclear quadrupole moment (Iodine has a nuclear spin of 5/2) is observed as splitting in the measured rovibrational profiles due to the hyperfine subcomponents, particularly for transition with $J \leq 2 \times K$. This agrees with the previous observations in the microwave region and in 11 μm regions [8, 31], and in the 3 μm region in the $\nu_4$ band [14]. The green ellipses in Fig. 5(b) show examples of the hyperfine subcomponents in the $^qP2(4)$ and $^qP2(3)$ transitions. Note that for transitions with $J > 2 \times K$, the hyperfine subcomponents show much smaller splitting, which cannot be resolved using Doppler-limited spectroscopy. In our spectral simulations, we excluded the hyperfine splitting because it is almost negligible for transitions with $J > 2 \times K$, representing the majority of assigned transitions.

Assignments of transitions in the complete spectral range from 2930 to 3010 cm$^{-1}$ were possible for $K$ from 0 to 13 with $J$ up to 88 in the fundamental $\nu_1/\nu_4$ level system, and for $K$ from 0 to 10 with $J$ up to 69 for the $\nu_3+\nu_1-\nu_3/\nu_3+\nu_4-\nu_3$ hot-band level system. Overall, 4665 transitions were assigned in the $\nu_1/\nu_4$ level system with an average error of 0.00071 cm$^{-1}$, of which 1753 transitions belong to the $\nu_1$ band and 2912 to the $\nu_4$ band. Note that 2063 transitions were assigned in the previous simulations of the $\nu_4$ band [14]. In the simulation of $\nu_1/\nu_4$ level system, in addition to the new line list of the $\nu_1$ band, we increased the number of assigned transitions of the $\nu_4$ band. For the hot-band level system, a total of 1239 transitions were assigned with an average error of 0.0045 cm$^{-1}$, of which 515 transitions were assigned to the $\nu_3+\nu_1-\nu_3$ band. A line list of all transitions assigned in this work is provided in the supplementary materials (S1).

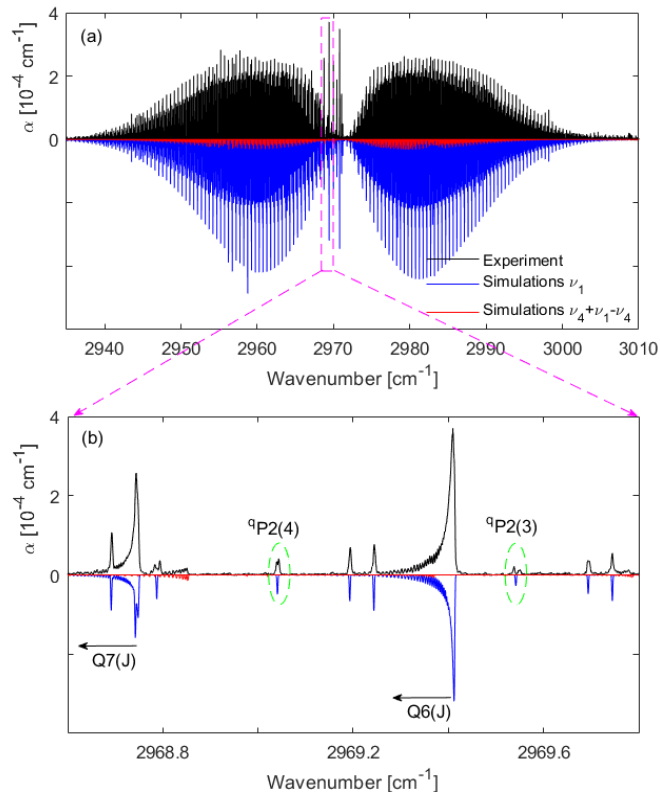

Fig. 5. (a) Measured absorption coefficient, α of 30 μbar of pure $CH_3I$ (black) at 296 K together with the simulations of the $\nu_1$ band (blue) and the $\nu_3+\nu_1-\nu_3$ hot band (red). The simulations are inverted for clarity. (b) Zoom in around Q6(J) – Q7(J) sub-band region. The arrows indicate the degradation direction within Q(J) with increasing $J$. The transitions $^qP2(4)$ and the $^qP2(3)$, marked with green ellipses, are examples of transitions with J ≤ 2 ×K that show signs of hyperfine splitting.

    The band parameters are obtained from the measured line positions by applying a fit with constrained lower state constants. The simulated spectrum is fitted to the measured spectrum and the nonlinear least squares analysis is performed to determine the corrections to the initial band parameters.

    Table 1 summarizes the parameters of the fundamental $\nu_1/\nu_4$ level system obtained in this work together with the band parameters reported by Paso et al. [7] and Anttila et al. [6] using an FT-IR spectrometer with 162 MHz resolution. The corresponding average error and the number of assigned transitions in the fitting pool are also included. As shown in this table, a difference in band origin of ~0.02 $cm^{-1}$ is observed for the $\nu_1$ band between our fit and that of Paso et al. [7], where they have treated the $\nu_1$ band separately without including assigned transitions for the $\nu_4$ in the fitting pool. For the $\nu_4$ band, a small difference in the band origin of $1.4 \times 10^{-4}$ $cm^{-1}$ is observed when it is modelled within the $\nu_1/\nu_4$ level system, as presented here, compared to when it is modelled separately, as in Ref. [14]. Table 2 shows a corresponding comparison for the $\nu_3+\nu_1-\nu_3$ and $\nu_3+\nu_4-\nu_3$ hot bands. Here, the connected four-level system treatment (current study) and the separate levels treatment (previous studies [6, 7, 14]) provide band origins with a difference of ~0.02 $cm^{-1}$ for the two hot bands. Nevertheless, the current spectroscopic model is considered more accurate because it improves the fit residuals globally as seen from the average errors in Tables 1 and 2, particularly for the $\nu_1$ band. Hence it provides more accurate band parameters of all the connecting states involved in the level systems.

    Tables 3 and 4 summarize the band parameters of the perturbing states as depicted in Fig. 3 in comparison with values reported by Paso et al. [7] and Anttila et al. [6]. The interaction parameters are listed in Table 5, together with those reported in earlier studies. The interaction parameters used in the hot-band level system were fixed to their values obtained from the fundamental band level system.

Table 1 Parameters of the $\nu_1$ and $\nu_4$ bands of $^{12}$CH$_3$I obtained from this study compared to previously reported parameters. Values in parentheses represent the fit parameter uncertainties at 1 σ confidence level.

| Constants [cm$^{-1}$] | $\nu_1$ band | | $\nu_4$ band | | |
|---|---|---|---|---|---|
| | Current work | Paso et al. [7] | Current work | Previous work [14] | Anttila et al. [6] |
| $\nu_0$ | 2971.81752(5) | 2971.8375(1) | 3060.07832(3) | 3060.07846(3) | 3060.07890(6) |
| A | 5.120586(2) | 5.12119(4) | 5.143968(2) | 5.144014(2) | 5.144038(4) |
| B | 0.25014258(5) | 0.25017(9) | 0.25033971(3) | 0.25033919(3) | 0.25034228(7) |
| $D_J \times 10^7$ | 2.11429(12) | 2.25(13) | 2.10005(7) | 2.09373(9) | 2.09410(14) |
| $D_{JK} \times 10^6$ | 3.1188(4) | 3.255(3) | 3.2673(6) | 3.3039(5) | 3.3139(17) |
| $D_K \times 10^5$ | 8.459(1) | 8.81(3) | 8.942(2) | 9.238(2) | 9.271(7) |
| $\zeta$ | | | 0.066351(11) | 0.066352(7) | 0.06655(10) |
| $\eta_J \times 10^7$ | | | -6.28(5) | -5.70(3) | -5.28(8) |
| $\eta_K \times 10^5$ | | | 3.10(3) | 4.33(2) | 4.43(7) |
| $q_+ \times 10^5$ | | | -1.0216(5) | -1.0102(4) | -1.052(5) |
| Number of lines | 1753 | 1600 | 2912 | 2063 | 1850 |
| Average error | 0.00071[a] | 0.0015 | 0.00071[a] | 0.00035 | 0.00083 |

[a] overall average error for the 4665 assigned transitions for the $\nu_1/\nu_4$ level system.

Table 2 Parameters of the $\nu_3+\nu_1-\nu_3$ and $\nu_3+\nu_4-\nu_3$ hot bands of $^{12}$CH$_3$I obtained from this study compared to previously reported parameters. Values in parentheses represent the fit parameter uncertainties at 1 σ confidence level.

| Constants [cm$^{-1}$] | $\nu_3+\nu_1-\nu_3$ band | | $\nu_3+\nu_4-\nu_3$ | | |
|---|---|---|---|---|---|
| | Current work | Paso et al. [7] | Current work | Previous work [14] | Anttila et al. [6] |
| $\nu_0$ | 2971.49736(2) | 2971.5240(7) | 3062.0247(4) | 3062.04686(5) | 3062.0474(2) |
| A | 5.11600(3) | 5.120(4) | 5.13785(2) | 5.147074(9) | 5.14704(4) |
| B | 0.2479640(7) | 0.2480(18) | 0.2485348(8) | 0.24852518(2) | 0.24852(17) |
| $D_J \times 10^7$ | 1.9285(2) | | 2.113(3) | 2.1015(6) | 2.09410 |
| $D_{JK} \times 10^6$ | 1.08(2) | | 3.03(2) | 2.886(6) | 3.0933(12) |
| $D_K \times 10^5$ | 9.8(4) | | 10.86(13) | 7.643(3) | 7.830(5) |
| $\zeta$ | | | 0.06701(2) | 0.068644(4) | .06865(6) |
| $\eta_J \times 10^6$ | | | -5.52(60) | -1.25(3) | -0.526 |
| $\eta_K \times 10^5$ | | | 9.2(11) | 2.85(2) | 4.43 |
| $q_+ \times 10^6$ | | | -4.6(8) | -9.12(2) | -9.44(5) |
| Number of lines | 515 | 500 | 724 | 831 | 380 |
| Average error | 0.0045[a] | 0.0070 | 0.0045[a] | 0.00084 | 0.0013 |

[a] overall average error for the 1239 assigned transitions for the $\nu_3+\nu_1-\nu_3/\nu_3+\nu_4-\nu_3$ level system.

Table 3 Parameters of the $\nu_2+2\nu_6^{\pm2}$ and $2\nu_2+\nu_3$ perturbing states in the fundamental band system obtained from this study compared to previously reported parameters. Values in parentheses represent the fit parameter uncertainties at 1 σ confidence level. Previously reported parameters without uncertainties were constrained in the fit.

| Constants [cm$^{-1}$] | $\nu_2+2\nu_6^{\pm2}$ | | | $2\nu_2+\nu_3$ | | |
|---|---|---|---|---|---|---|
| | This work | Paso et al. [7] | Anttila et al. [6] | This work | Paso et al. [7] | Anttila et al. [6] |
| $\nu_0$ | 3017.9859(4) | 3017.819(2) | 3017.992(14) | 3010.721(7) | 3009.5 | 3010.79(3) |
| A | 5.25091(5) | 5.2606 | 5.2606(3) | 5.2636(2) | 5.2071 | 5.2606 |
| B | 0.247226(1) | 0.2471 | 0.2471(2) | 0.24839(2) | 0.24569(2) | 0.24702 |
| $D_J \times 10^7$ | 1.938(2) | | | 6.0192(2) | | |
| $D_{JK} \times 10^6$ | 8.17(2) | | | 6.056(3) | | |
| $D_K \times 10^5$ | -14.7(7) | | | 13.2(8) | | |
| $\zeta$ | -0.3913(2) | -0.385 | -0.385(14) | | | |
| $\eta_J \times 10^6$ | -10.80(8) | -13.43 | -13.43 | | | |
| $\eta_K \times 10^4$ | -15.30(4) | -2.87 | -2.87 | | | |

Table 4 Parameters of the $\nu_2+\nu_3+2\nu_6^{\pm2}$ and $2\nu_2+2\nu_3$ perturbing states in the hot bands obtained from this study compared to previously reported parameters. Values in parentheses represent the fit parameter uncertainties at 1 σ confidence level. Previously reported parameters without uncertainties were constrained in the fit.

| Constants [cm$^{-1}$] | $\nu_2+\nu_3+2\nu_6^{\pm2}$ | | $2\nu_2+2\nu_3$ | |
|---|---|---|---|---|
| | This work | Paso et al. [6, 7] | This work | Paso et al. [6, 7] |
| $\nu_0$ | 3005.847(14) | 3007.70(4) | 2993.48(12) | 2992.79 |
| A | 5.2362(4) | 5.2543 | 5.2644(10) | 5.2007 |
| B | 0.24712(1) | 0.24528 | 0.24809(2) | 2.4284(2) |
| $\zeta$ | -0.390(7) | -0.386 | | |

Table 5 Interaction parameters obtained in the comb-based FTS measurements compared to those from previous lower resolution FT-IR measurement Values in parentheses represent the fit parameter uncertainties at 1 σ confidence level. Previously reported parameters without uncertainties were constrained in the fit.

| Constants [cm$^{-1}$] | This work | Paso et al [7] | Anttila et al [6] | Alanko et al. [22] [a] |
|---|---|---|---|---|
| $C_{4266} \times 10^3$ | 13.21(6) | | 12.964(7) | 12.27(11) |
| $C^J_{4266} \times 10^7$ | 4.51(2) | | 4.82(3) | 6.52(62) |
| $F_{4266}$ | 0.742(2) | | 0.7367(4) | 0.8149(7) |
| $F^J_{4266} \times 10^6$ | -9.2(1) | | -12.5(4) | -1.85(90) |
| $C_{4223} \times 10^3$ | 2.9(9) | | | |
| $C^J_{4223} \times 10^7$ | -4.36(40) | | | |
| $C_{14} \times 10^3$ | 1.7(2) | 1.10(16) | | 4.05(33) |
| $C^J_{14} \times 10^7$ | -7.06(60) | | | |
| $C_{1266} \times 10^3$ | 9.1(1) | 9.29(3) | | |
| $C^J_{1266} \times 10^7$ | -2.94(3) | -4.18(9) | | |
| $F_{1223}$ | 4.748(1) | 4.75 | | 3.360(30) |
| $F^J_{1223} \times 10^5$ | 1.10(5) | -6.6(5) | | 32.9(12) |

[a]: for the $^{13}CH_3I$

Figure 6 shows the residuals of the least squares fit to the 4655 transitions assigned in the $\nu_1/\nu_4$ level system as a function of J and K quantum numbers [panel (a)], and the 1239 transitions assigned in the $\nu_3+\nu_1-\nu_3/\nu_3+\nu_4-\nu_3$ level system [panel (b)]. Both panels show residuals randomly scattered around zero for transitions with quantum numbers up to $J = 88$ and $K = 13$ for the $\nu_1/\nu_4$ system and up to $J = 69$ and $K = 10$ for the $\nu_3+\nu_1-\nu_3/\nu_3+\nu_4-\nu_3$ level system. The average error of 0.00071 cm$^{-1}$ reported for the $\nu_1/\nu_4$ level system is smaller than the 0.0015 cm$^{-1}$ error previously reported for the $\nu_1$ band by Paso et al [7], and the 0.00085 cm$^{-1}$ error for the $\nu_4$ band reported by Anttila et al. [6]. We note that the average error is a factor of two larger than that reported for the $\nu_4$ band in our previous work [14], which we attribute to the large scatter of the assigned transitions of the $\nu_1$ band, particularly for higher K values, as can be seen in Fig. 6(a).

Our simulations show a good global agreement with the measurements, and a smaller average error for a larger number of assigned transitions than the earlier FT-IR measurements, and hence our reported band parameters are considered more accurate. This accuracy is enabled by the high spectral resolution of the measured spectrum [14], and the spectroscopic treatments of the $\nu_1/\nu_4$ and $\nu_3+\nu_1-\nu_3/\nu_3+\nu_4-\nu_3$, each as interacting four-level system with perturbing states. Please note that the slightly larger residuals for higher K numbers of the $\nu_1$ band, see Fig. 6(a), may suggest the existence of another connection to one of the vibrational bands in the parallel vibration region, extending from 2800 to 2900 cm$^{-1}$ [10], most likely the $2\nu_5$ band as suggested by Alanko et al. [22] in his analysis of the $^{13}CH_3I$ rovibrational spectrum in the mid-IR region. Therefore, a new assignment of the parallel vibration region utilizing the high-precision frequency comb measurements and investigating the possible global perturbations to the $\nu_1/\nu_4$ level systems will be commenced in future studies.

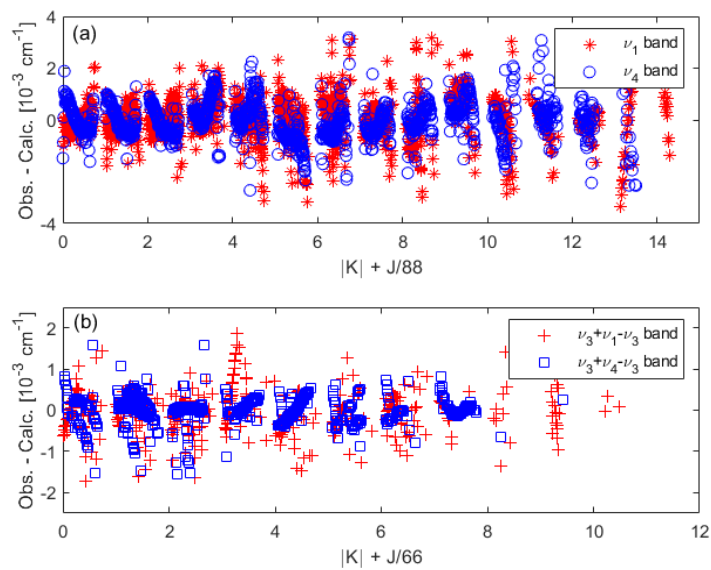

Fig. 6. Fit residuals (Obs. – Calc.) as a function of the upper state K and J quantum numbers of the (a) $\nu_1/\nu_4$ band system, and the (b) $\nu_3+\nu_1-\nu_3/\nu_3+\nu_4-\nu_3$ bands.

### 3.3. Line intensities

Line intensities of selected assigned lines of the $\nu_1$ band and the $\nu_3+\nu_1-\nu_3$ hot band were obtained using a multispectral fitting routine based on the Levenberg-Marquardt algorithm applied to spectra measured at 0.11 mbar, 0.15 mbar, 0.20 mbar and 0.25 mbar of $CH_3I$ diluted in dry air to a total pressure of 10 mbar. We fitted lines in windows of ±4 FWHM, where the width was calculated as the quadrature sum of the mean Doppler width and the mean Lorentzian component. We fitted lines separated by less than 6 times the FWHM in a common window. We allowed a maximum of three lines per fit window and did not fit larger clusters. Furthermore, the minimum allowed separation for lines to be fitted was 2 times the FWHM.

We made an initial identification of absorption lines in the spectrum measured at 0.11 mbar of pure $CH_3I$ using a MATLAB peak finding tool and designated fitting windows subject to the criteria defined above. We set a minimum absorption threshold to $1\times10^{-5}$ cm$^{-1}$, and also a maximum threshold of $1.8\times10^{-4}$ cm$^{-1}$ since the $\nu_1$ band contained a large number of saturated lines. We selected the maximum threshold in order to have at least one unsaturated data set for the multispectral fit. Next, for accurate determination of the line center frequencies, we fit the lines using the Voigt model with the line center and intensity as free parameters. The Lorentzian FWHM was fixed to a value of 28 MHz, which was attributed to instrumental broadening [14], and the Doppler width was fixed to the theoretical value at each line position, around 92 MHz FWHM. The fitted lines were further compared with the spectrum simulated in PGOPHER, to determine the lines that could be unambiguously assigned. We then applied the multispectral fit to the four measurements at 10 mbar of total pressure, in all fit windows containing at least one assigned line. The fit to the 0.11 mbar pure $CH_3I$ provided the initial parameters for the multispectral fit, and the free parameters were again the line position and the line intensity, with the Lorentzian broadening fixed to the same global optimum found previously for the $\nu_4$ band in Ref. [14]. For each fit window, measurements in which the strongest line exceeded a peak absorption of $1.2\times10^{-4}$ cm$^{-1}$ were excluded to avoid saturation effects. We rejected fits for which the quality factor was lower than 5 at the lowest pressure, with the quality factor defined as the peak absorption divided by the standard deviation of the residuals in a region of ± 4 FWHM around the line center. We inspected the remaining fits visually to eliminate those affected by baseline issues caused by nearby saturated lines or showing signs of overlapping weak lines. Fig. 7 shows an example of the fitted absorption profiles at different partial pressures of $CH_3I$ together with the model for $^qP4(64)$ and $^qP2(65)$ transitions around 2938.094 cm$^{-1}$ and 2938.106 cm$^{-1}$, respectively as well as for the degenerate $^qP6(62)$ transitions around 2938.136 cm$^{-1}$. The lower panel shows the fit residuals.

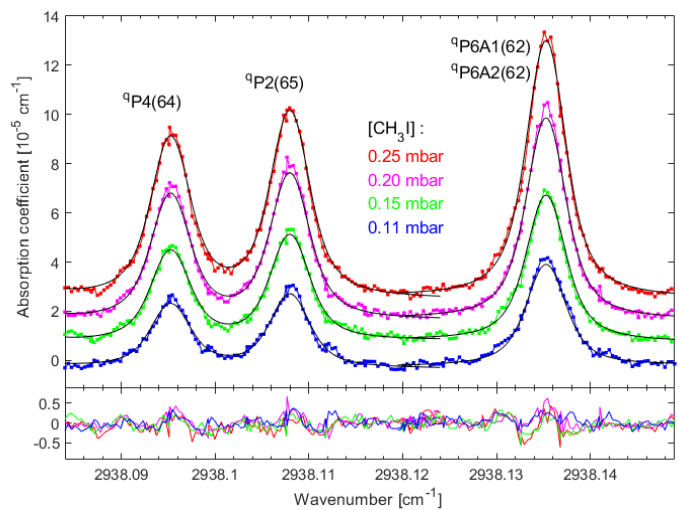

Fig. 7. Absorption coefficients of selected transitions measured at 0.11 mbar (blue), 0.15 mbar (green), 0.20 mbar (pink), and 0.25 mbar (red) of $CH_3I$ at total pressure of 10 mbar in dry air (vertically offset for clarity) together with the multispectral fit (black line). The lower panel shows residuals.

Finally, we report the spectral line intensities for a selected set of 49 absorption profiles of the $\nu_1$ band belonging to transitions up to $K$=12, distributed over the range covered by the $\nu_1$ band, and 4 profiles of the $\nu_3+\nu_1-\nu_3$ band. Note that a number of these profiles are assigned to degenerate line pairs of $A_1/A_2$ symmetry for which the fitted intensities and corresponding uncertainties were distributed equally. The final line list hence includes 65 transitions of the $\nu_1$ band and 7 of the $\nu_3+\nu_1-\nu_3$ band, and their intensities are provided in supplementary material (S2), together with their assignments as well as line positions obtained from the Voigt fits to the measurement at 0.11 mbar pure $CH_3I$. Summing in quadrature the estimated uncertainties due to instrumental broadening (5%), the fit (0.6% on average), the pressure measurements (4%), the multipass cell path length (1.3%), and the sample purity (1%) yields a 7% relative uncertainty in the line intensities. Considering other possible error sources that may exist, e.g., overlapping weak transitions from the fundamental or hot bands, we conservatively estimate 10% as a total uncertainty in the reported intensities.

## 4. Conclusions

We used the high-resolution mid-infrared spectrum of $CH_3I$ measured using comb-based Fourier transform spectroscopy in the range from 2900 cm$^{-1}$ to 3160 cm$^{-1}$ [14] to extend the spectral analysis to the $\nu_1$ and the $\nu_3+\nu_1-\nu_3$ bands around 2971 cm$^{-1}$. The current treatment models the two fundamental bands, $\nu_1$ and $\nu_4$, which are close in energy, together with the perturbing states, $2\nu_2+\nu_3$ and the $\nu_2+2\nu_6^{\pm2}$ as a four-level system, interacting through a number of Fermi and Coriolis interactions. The two hot bands, $\nu_3+\nu_1-\nu_3$ and $\nu_3+\nu_4-\nu_3$, are also assumed to form a four-level system connected with the perturbing states $2\nu_2+2\nu_3$ and $\nu_2+\nu_3+2\nu_6^{\pm2}$. This treatment yields a good global agreement between the simulated and the measured spectra, and hence more accurate band parameters. In addition, it allows us to revisit our previous assignment of the $\nu_4$ and the $\nu_3+\nu_4-\nu_3$ bands, where we treated them separately and not in connection with their $\nu_1$ and the $\nu_3+\nu_1-\nu_3$ counterparts as done in the current work.

Overall, we assigned 4655 transitions in the $\nu_1/\nu_4$ level system, and the 1239 transitions in the $\nu_3+\nu_1-\nu_3/\nu_3+\nu_4-\nu_3$ level system. The average error of the least squares fit to the assigned transitions in the $\nu_1/\nu_4$ level systems is 0.00071 cm$^{-1}$, which is smaller than those previously reported for the $\nu_1$ band of 0.0015 cm$^{-1}$ [7], and for the $\nu_4$ band of 0.00085 cm$^{-1}$ [6]. For the hot band system an average error of 0.0045 cm$^{-1}$ was obtained from the fit to the assigned transitions.

The $\nu_1$ band shows hyperfine splitting, resolvable for transitions with $J \leq 2 \times K$. The observation of these fine structures was not possible in the earlier measurements of Paso et al. [7], using FT-IR with a resolution of almost two times the Doppler width of $CH_3I$. We also report the spectral intensities of 65 lines of the $\nu_1$ band and 7 lines of the $\nu_3+\nu_1-\nu_3$ band for the first time using the Voigt line shape as a model in multispectral fitting. The reported line lists and intensities will serve as a basis for developing a semi-empirical spectroscopic model for high-resolution molecular spectroscopic databases, e.g., HITRAN [15] and GEISA [32]. It will also serve as a basis for line selection in future monitoring applications of $CH_3I$ using sensitive spectroscopic techniques [33].

**Acknowledgements**

The authors thank the late Colin Western for providing help with simulating the spectrum in PGOPHER. This project is financed by the Knut and Alice Wallenberg Foundation (KAW 2015.0159 and 2020.0303) and the Swedish Research Council (2016-03593, 2020-00238). I. Sadiek would like to thank the German Research Foundation for financing his current position at the Leibniz Institute for Plasma Science and Technology (SA 4483/1-1).

**Supplementary materials**

Supplementary material associated with this preprint can be obtained from the corresponding author.